\documentclass[a4paper,fleqn]{cas-sc}

\usepackage[authoryear]{natbib}
\usepackage{amsfonts,amstext,amssymb,amsmath}
\usepackage{calrsfs}
\usepackage[svgnames,dvipsnames]{xcolor}
\usepackage{multicol}
\usepackage{subfig}
\usepackage{graphicx}
\usepackage{nicefrac}
\usepackage{physics}
\usepackage{todonotes}
\usepackage{ulem}
\usepackage{soul}

\usepackage{tabularx}
\usepackage{booktabs}
  
\usepackage{hyperref}
\usepackage{wrapfig}

\usepackage[noabbrev,capitalise,nameinlink]{cleveref}
\hypersetup{colorlinks={true},linkcolor={blue},citecolor=blue}

\DeclareMathAlphabet{\pazocal}{OMS}{zplm}{m}{n}

\def\tsc#1{\csdef{#1}{\textsc{\lowercase{#1}}\xspace}}
\tsc{WGM}
\tsc{QE}
\tsc{EP}
\tsc{PMS}
\tsc{BEC}
\tsc{DE}

\usepackage{todonotes}

\begin{document}
\let\WriteBookmarks\relax
\def\floatpagepagefraction{1}
\def\textpagefraction{.001}

\shorttitle{Designing necks and wrinkles in inflated auxetic membranes}

\shortauthors{Sairam Pamulaparthi Venkata et~al.}

\title [mode = title]{Designing necks and wrinkles in inflated auxetic membranes}                      
\tnotemark[1]

\tnotetext[1]{This research project funded by the XS-META ITN: Marie Sk{\l}odowska-Curie European Actions (Grant Agreement No. 956401).}


%
\author[1]{Sairam Pamulaparthi Venkata}[type=,
                        auid=000,bioid=1,
                        prefix=,
                        role=,
                        style = english,
                        orcid=0000-0002-2409-7332]


\ead{S.PamulaparthiVenkata1@universityofgalway.ie}


\credit{Writing, methodology, numerical simulations}

\affiliation[1]{organization={School of Mathematical and Statistical Sciences, University of Galway},
    addressline={University Road}, 
    city={Galway},
    postcode={H91 TK33}, 
    country={Ireland}}

\author[1]{Valentina Balbi}[
   orcid=0000-0002-7538-9490
   ]
\ead{vbalbi@universityofgalway.ie}

\credit{Conceptualisation, writing}

\author[1,2]{Michel Destrade}[style=english, orcid=0000-0002-6266-1221]
\ead{michel.destrade@universityofgalway.ie}

\affiliation[2]{organization={Key Laboratory of Soft Machines and Smart Devices of Zhejiang Province and Department of Engineering Mechanics, Zhejiang University},
    postcode={Hangzhou 310027}, 
    country={People’s Republic of China}}

\credit{Conceptualisation, writing}
  
\author[1]{Giuseppe Zurlo}[style=english,orcid=0000-0002-1438-5015]
\cormark[1]
\ead{giuseppe.zurlo@universityofgalway.ie}
\credit{Conceptualisation, methodology, writing}
\cortext[cor1]{Corresponding author}

\begin{abstract}
Thin elastic membranes may undergo various types of instabilities, depending on material properties, mechanical loading and boundary conditions. In this work, we focus on axially symmetric and functionally graded membranes, and we study their behavior upon inflation of an initially flat disk. 
Our findings reveal that when the membrane is auxetic, its material properties can be functionally graded to produce wrinkles and necks at desired locations upon inflation. 

To overcome the complexities arising from strong geometric and constitutive nonlinearities, we use a semi-analytical approach to identify the right functional grading required to produce nontrivial instabilities. 
Hence we use the method of tension field theory to describe the onset of wrinkling in an averaged way. We obtain a series of universal results providing powerful insights into the formation of wrinkles and necks in inflated membranes. For example, we prove analytically that necks and wrinkles may never overlap in inflated, axially symmetric membranes.  

Guided by these universal results, we implement the relaxed strain energy method into a Finite Element solver (\texttt{COMSOL}). By tuning spatial inhomogeneities of the material moduli, we are able to generate specific instability patterns at desired locations. In particular, we obtain necks alone, wrinkles alone, or alternating necking and wrinkling patterns.
Moreover, we are able to delay or anticipate the snap-through behavior of the inflated membrane.

These results will help with the deployment of functionally-graded auxetic membranes for applications where inflated or laterally pulled membranes may be morphed and corrugated on demand. 
\end{abstract}




\maketitle



\section{Introduction\label{intro}}


Artificially designed materials with negative Poisson's ratio thicken when stretched, in contrast to classical elastic materials, which become thinner in the directions lateral to an applied loading direction. 
These so-called auxetic materials have been studied in several fields, including ferromagnetics \citep{popereka1970ferromagnetic}, crystal elasticity \citep{milstein1979existence}, foam structures \citep{lakes1987foam}, microporous materials \citep{evans1989microporous}, and composites \citep{milton1992composite}.

Theoretically, for isotropic materials which satisfy the pointwise energy stability criterion, the Poisson ratio can take values between -1 and 0.5 for 3D solids \citep{love1944treatise, timoshenko1983history}. For 2D solids such as membranes, the allowable Poisson's ratio value lies between -1 and 1 \citep{wojciechowski1989negative}.
In contrast, Poisson's ratio for anisotropic structures has no bounds \citep{Ting2005}, which is why some biological tissues exhibit auxeticity \citep{frolich1994poisson,bowick2001universal, lakes2001broader}.

\begin{figure*}[!h]
\centering
\includegraphics[scale=0.5]{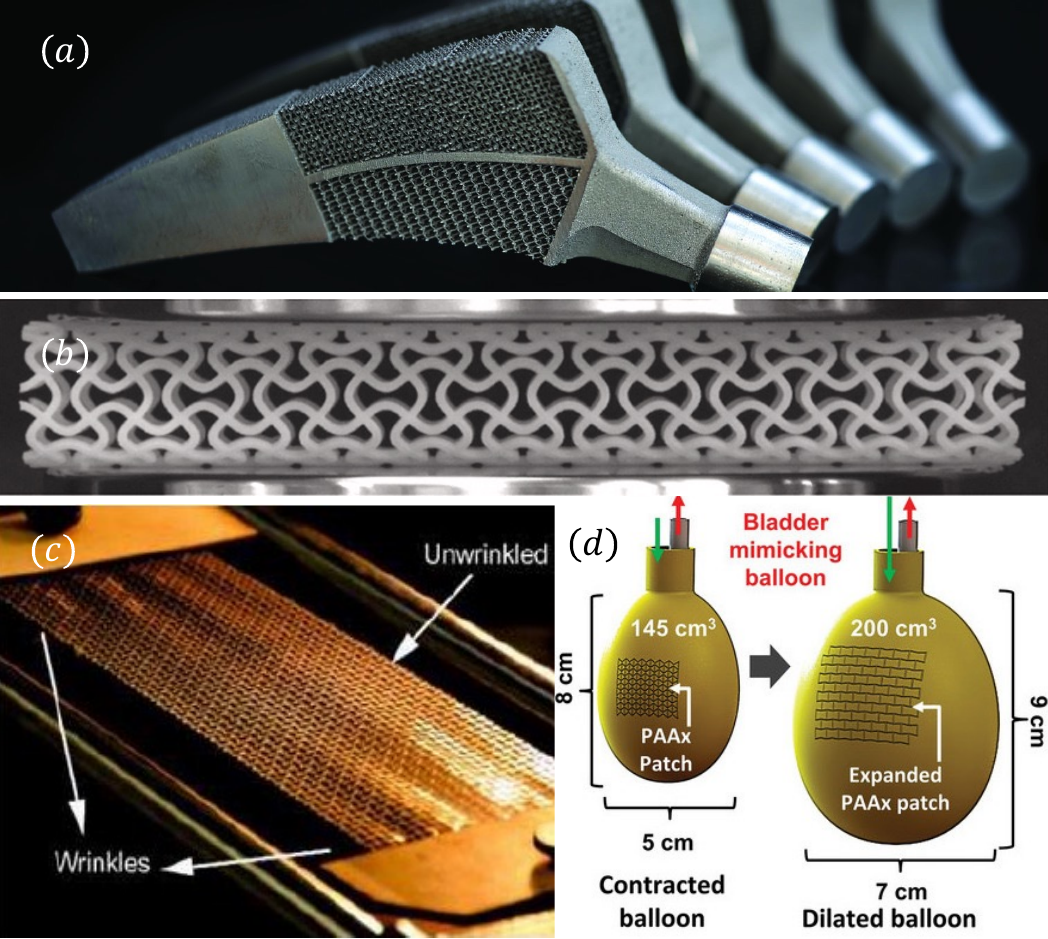}
    \caption{Examples of applications and behaviors for auxetic materials. (a) Hip implants can be designed to include regions with a negative Poisson ratio to minimize retraction from the bone under biomechanical loading \citep{kolken2018rationally}.
    (b) Auxetic tubular lattice stents exhibit increased ductility compared to conventional diamond tubular lattices \citep{jiang20223d}.
    (c) Auxetic thin membranes under tension wrinkle in the neighborhood of clamps, in contrast to conventional membranes, which wrinkle in their central region \citep{bonfanti2019elastic}.
    (d) Auxetic patches glued onto balloon membranes can undergo larger deformations than conventional patches, a property which can help with the healing of puncture wounds to the bladder for example \citep{Nguyen2022}. 
    }
    \label{fig:applications}
\end{figure*}

Auxetic materials can be fabricated using additive manufacturing techniques, such as powder bed fusion \citep{king2015laser, sun2017powder} or subtractive manufacturing techniques, such as laser cutting for thin structures \citep{Mueller2013, BHULLAR2017334}. With the rapid advancement in additive manufacturing techniques, even complex auxetic structures can be produced at large scales in a short time. With a careful design of voids or holes at the micro-scale, auxetic properties can be obtained at the continuum level \citep{Lakes1993, bertoldi2017flexible}. 

These structures have a wide range of applications, from soft robotics, with compliant actuators \citep{lazarus2015soft} or compliant grippers \citep{kaur2019toward}, to biomedical applications, with auxetic stents \citep{Dolla2006}, dynamic organ patches \citep{Nguyen2022}, or skin grafts \citep{gupta2022biomechanical}.
\cref{fig:applications} displays examples of recent applications.

Due to their thin-walled, lightweight, and impressive tensile properties, nonlinear elastic membranes play a prominent role in many fields such as automobile, aerospace, civil and biomedical engineering \citep{beatty1987topics, evans2009implementation, fu2016solar}. However, thin membranes lose their mechanical stability under in-plane compressive stresses due to negligible bending rigidity, leading to interesting behaviors \citep{timoshenko1983history, roddeman1987wrinkling, cerda2002wrinkling}. Inflatable membranes under large elastic deformations experience various kinds of bifurcation phenomena such as limit-point (snap-through), wrinkling, and necking instabilities. Some of the most common and significant applications of inflatable auxetic materials can be seen in stent deployment for angioplasty \citep{amin2015auxetic}, in morphing structures \citep{sun2014active}, and for smart biomaterials \citep{bhullar2015smart}.

 Over the past several decades, there has been extensive research on finite inflation in incompressible circular membranes. Early experimental works by \cite{flint1937physical}, \cite{treloar1944strains}, and \cite{rivlin1951large} present a detailed analysis of the deformation shape and strain distributions over the surface of the inflated incompressible isotropic balloons. 
 \cite{adkins1952large} studied theoretically pressurized circular and spherical thin shells using incompressible neo-Hookean and Mooney-Rivlin strain energy functions, to obtain the relationship between inflating pressure, extension ratio, and radius of curvature near the pole. 
The book by \cite{green1960large} covers the mathematical theory of finite strain deformations in nonlinear membranes extensively for both isotropic and anisotropic materials. 
More recent experimental works include that by \cite{machado2012membrane}, who proposed a method to determine curvatures and membrane stresses using an axisymmetric bulge test for isotropic circular membranes and a three-dimensional digital image correlation technique. 
 
 
 Finite element analysis (FEA) is very useful in understanding the behavior of complex geometries and loading conditions. Many researchers have developed numerical methods based on FEA to study finite inflation in nonlinear membranes using different material models and for different geometries. For example, see the works on axisymmetric membranes by \cite{oden1967finite}, \cite{wriggers1990fully}, \cite{gruttmann1992theory}, \cite{jiang1995finite}, \cite{rumpel2005efficient}, \cite{eriksson2012instability}, and \cite{selvadurai2022mechanics}, on rectangular membranes using finite difference iterative scheme by \citet{yang1973general}, and on square membranes using FEA by \citet{Aaron2001}, \citet{lee2006finite}, \citet{barsotti2014static}, and \citet{CHEN2014}. 
 
 Analytical solutions associated with the finite inflation of nonlinear membranes are scarce in the literature due to strong material and geometrical nonlinearities. 
 However, analytical solutions play an important role in providing simplified and direct solutions to predict the mechanical behavior of inflated membranes. 
 With an assumption of linear elastic constitutive behavior and spherical deformation shape for a pressurized incompressible isotropic circular membrane, \cite{fichter1997some} provided closed-form analytical solutions for small deformations, later extended by \cite{coelho2014numerical} for finite strains. 
 Relaxing the constraint of linear elastic material behavior but still assuming the spherical deformation shape of the membrane, \cite{yuan2021analytical}, \cite{foster1967}, and \cite{yang2021contact} derived analytical solutions for pre-stretched circular membranes under inflation using different incompressible hyperelastic material models. Dropping the hypothesis of spherical deformation shape and considering compressibility, pressure-deflection formulae for inflated isotropic circular membranes with compressible Mooney-Rivlin material model are provided by \cite{pelliciari2022analytical} without pre-stretch,  by \cite{sirotti2023analytical} with pre-stretch, and by \cite{pelliciari2022continuum} for anisotropic pressurized graphene membranes without prestretch.

Many of the above-mentioned works dealt with limit-point instability in inflatable circular membranes, but little attention has been drawn in the past to the study of wrinkling and necking instabilities for functionally-graded inflatable auxetic circular membranes under mechanical loads.

Several theories exist in the literature for studying wrinkles in nonlinear elastic membranes. 
For example, the theory of incremental deformations \citep{haughton1978incr, haughton1978increm}, the Föppl–von Kármán theory of plates \citep{dym1973solid, puntel2011wrinkling}, the reduced-order finite element membrane theory \citep{damil2010influence}, and the numerical bifurcation-continuation analysis \citep{healey2013wrinkling}. Although these advanced and refined models provide comprehensive details on the wavelength and amplitude of wrinkles, they are computationally expensive to implement. 

 In this study, we focus on determining the average deformation in the wrinkled region along with finding the location and orientation of wrinkles, but not their amplitude and wavelength. Therefore, we employ tension field theory, originally proposed by \cite{wagner1931flat} and \cite{reissner1938tension}. Tension field theory has the advantage of being computationally more viable and mathematically elegant. According to tension field theory, membranes are assumed to have zero out-of-plane bending stiffness and cannot sustain in-plane compressive stresses. 
 In the 1980s, in order to account for compressive stresses, \cite{Pipkin1986TheRE} extended the theory by introducing the concept of ``relaxed strain energy function'', see also \cite{pipkin1986continuously}, \cite{Steigmann1989wrinkling}, \cite{steigmann1990tension}, and \cite{Pipkin1994}. Relaxed models have been used to study wrinkles in anisotropic membranes \citep{PipkinAniso}, electroactive elastomeric membranes \citep{DeTommasi2011,DeTommasi2012,GREANEY201984,KhuranaA,KhuranaB}, pressurized magnetoelastic circular membranes \citep{SAXENA2019250}, and inflatable isotropic membranes under uniform pressure load \citep{tamadapu2014, Riccardo2015, pamulaparthi2019instabilities}. 

Alongside wrinkling, necking is another interesting bifurcation phenomenon observed in nonlinear membranes, although less explored in hyperelastic isotropic circular membranes. \cite{chaudhuri2014static} found negative Gaussian curvature and circumferential wrinkling at the fixed rim of the hyperelastic isotropic circular membrane under inflation. Necking in pressurized elasto-plastic spherical membranes has been investigated by \cite{needleman1976necking} and in pressurized elasto-plastic circular membranes by \cite{chater1983}. However, a study on necking and multi-layered bubbling phenomenon in pressurized functionally-graded hyperelastic isotropic auxetic circular membranes is still missing in the literature.

In this work, as a proof-of-a-concept, we study the effect of varying material properties such as the Young modulus and the Poisson ratio on limit-point instabilities (snap-through), necking, and wrinkling in pressurized isotropic auxetic membranes (circular and square geometries). We use the Blatz-Ko strain energy function to model the membrane's hyperelastic mechanical behavior. 
We study the effect of pre-stretches on 
wrinkling instabilities in auxetic circular membranes.
For the onset of limit-point instabilities, we investigate the effect of pre-stretches and material parameters for circular geometries.
Finally, we show how to obtain wrinkling patterns in specific areas of the membrane by tuning the spatial distribution of Young's modulus and Poisson's ratio, both for square and circular membranes. 

 
The paper is organized as follows. 
In \cref{matmodel}, we introduce the problems of interest and the constitutive equations, and we derive the kinematics of the deformation. 
We also briefly summarize the main features of the relaxed strain energy functional based on tension field theory to derive the membrane stresses. Finally, we write down the equilibrium conditions along with the applied boundary conditions. In \cref{insights}, with the help of membrane theory, we derive the equations linking curvatures and principal stresses, which are necessary for wrinkling and necking in the circular membrane. We establish several universal insights, valid for all hyperelastic isotropic membranes.
These include the results that regions of necking and wrinkling cannot overlap, and that necking and wrinkling cannot occur in the center of a pre-stretched circular membrane.  

In  \cref{results1}, we compare the results of our finite element simulations in \texttt{COMSOL} \citep{multiphysics2022} with the universal predictions of  \cref{insights}. Conclusions and limitations of the current work, along with possible directions for future works are detailed in \cref{conclusions}.  These include some preliminary results on square membranes which can be inflated to exhibit a desired wrinkling pattern. These results could be used in applications involving haptics.



\section{Membrane deformations, energy and stress\label{matmodel}}


In this work, we describe the behavior of inflated elastic membranes that are rotationally symmetric about an axis. In the so-called ``membrane approximation'', the membrane thickness is small in comparison to its diameter and bending effects are neglected. In this approximation, the 3D deformation is deduced from the 2D deformation of the membrane mid-surface.  

\begin{figure*}[thb]
\centering
\includegraphics[scale=0.9]{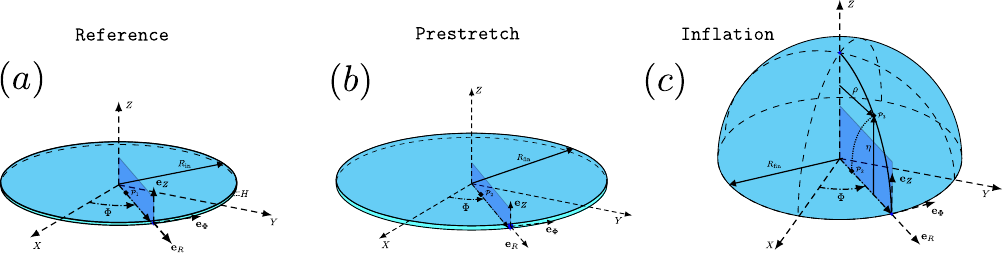}
    \caption{Deformation profiles of the axisymmetric auxetic circular membrane. 
    (a): Circular membrane with  initial radius of $R_{\text{in}}$ and thickness $H$. 
    (b): Deformed profile of the circular membrane under radial pre-stretch with the radius $R_{\text{fin}}$. 
    Note that the thickness of the membrane is increased with the in-plane stretching due to auxeticity. 
    (c): Deformed profile of the inflated circular membrane subjected to uniform pressure $P$ and fixed on its circumferential boundary. 
    Point $\pazocal{P}_{1}$ in the undeformed configuration is displaced to the positions $\pazocal{P}_{2}$ and $\pazocal{P}_{3}$ with pre-stretch and inflation, respectively. 
    A point in the final configuration has coordinates $(\rho, \Phi, \eta)$, where $\rho$ is radial position and $\eta$ is the vertical deflection.}
    \label{fig:cir-geom1}
\end{figure*}

We consider a circular membrane with radius $R_{\text{in}}$ and we use a cylindrical coordinate system to represent the kinematics of the deformation. We identify the position of a point on the mid-surface of the membrane in its undeformed configuration with $\pazocal{P}_{1}\left(R, \Phi, 0\right)$. 
As the membrane is radially stretched axisymmetrically, the point $\pazocal{P}_{1}$ moves to position $\pazocal{P}_{2} \left(\rho_{0}, \Phi, 0\right)$ in the membrane with radius $R_{\text{fin}}$. Upon axisymmetric inflation of the pre-stretched membrane (with a fixed circumference) under a uniform pressure $P$  from the side $Z=0^-$, the membrane bulges out of the plane towards $Z=0^+$, and the point $\pazocal{P}_{2}$ is displaced to position $\pazocal{P}_{3} \left(\rho(R), \Phi, \eta(R)\right)$. Here, $\rho(R)$ and $\eta(R)$ represent the radial and transverse deflections of the point $\pazocal{P}_{1}$ in the final configuration, respectively, as shown in \cref{fig:cir-geom1}.

The in-plane stretches of the 2D membrane are calculated as
\begin{equation}
\lambda_{1}  = \sqrt{\left(\dv{\rho}{R}\right)^2 + \left(\dv{\eta}{R}\right)^2}, \qquad \lambda_{2}  = \frac{\rho}{R},
\label{eq:prinstrch3}
\end{equation}
where the index $1$ refers to meridians and $2$ to parallels in the current configuration. 

In the membrane approximation, it is assumed the deformation is normal preserving. Therefore the gradient of deformation admits the diagonal representation $\mathbf{F}=\text{diag}(\lambda_{1} ,\lambda_{2} ,\lambda_{3})$ in a local basis, where the eigenvalue $\lambda_3$ is relative to the membrane normal direction. By introducing the volumetric variation coefficient $J = \det \mathbf{F}$, we may therefore write 
\begin{equation}
\qquad \lambda_{3} = \frac{J}{\lambda_{1}\lambda_{2}}. 
\end{equation}
With $\textbf{F}$ written this way, we may now compute the 2D membrane energy from any 3D energy. Specifically here, to model the compressible behavior of the membrane, we use the three-dimensional Blatz-Ko strain energy density \citep{blatz1962application, BROCKMAN1986}, 
\begin{equation}
W^{\text{3D}} = \alpha \frac{\mu}{2}\left[I_1-3+\frac{1-2\nu}{\nu}\left(I_3^{-\frac{\nu}{1-2\nu}}-1\right)\right]+(1-\alpha) \frac{\mu}{2}\left[\frac{I_2}{I_3} - 3 + \frac{1-2\nu}{\nu}\left(I_3^\frac{\nu}{1-2\nu}-1\right)\right],
\label{eq:energypart6a}
\end{equation}
where $I_1=\text{tr}(\mathbf{FF}^T)$, $I_3 = J^2$, $I_2 = I_3\text{tr}(\mathbf{FF}^T)^{-1}$, and 
$0<\alpha<1$, $\mu>0$, $-1 < \nu \le 1/2$ are material constants. 
The infinitesimal shear modulus $\mu$ and infinitesimal Poisson ratio $\nu$ are related through $\mu = \nicefrac{E}{2(1+\nu)}$, where $E$ is the infinitesimal Young modulus.

 With the help of the plane-stress state condition \citep{haughton1995wrinkling, haughton1996wrinkling}, $P_{33}^{\text{3D}}  = 0$, where $\mathbf{P}^{\text{3D}}  = \partial W^{\text{3D}} /{\partial \mathbf{F}}$ is the first Piola-Kirchhoff stress, we find the out-of-plane principal stretch ratio $\lambda_3$ as
\begin{equation}
\lambda_{3} = \left(\lambda_{1}\lambda_{2}\right)^{-\frac{\nu}{1-\nu}} = \Lambda^{\frac{1}{2}}, \qquad \text{ where } \qquad \Lambda = \left(\lambda_{1}\lambda_{2}\right)^{-\frac{2\nu}{1-\nu}}.
\label{eq:stretch3rd7}
\end{equation}
Now, by substituting \cref{eq:stretch3rd7} in \cref{eq:energypart6a}, we obtain $W$, i.e. the \textit{membrane strain energy function}, in terms of the two in-plane principal stretch ratios $\lambda_{1}$ and $\lambda_{2}$, as follows
\begin{equation}
W = E\frac{ \alpha\left(\lambda_1^2+\lambda_2^2\right) + \left(1-\alpha\right)\left(\lambda_1^{-2} + \lambda_2^{-2}\right) - 1}{4(1+\nu)} + E \frac{\left( \left(\alpha - 1\right)\Lambda^{-1} - \alpha\Lambda \right)\left(\nu - 1\right) - 1}{4\nu(1+\nu)}.
\label{eq:memenergy8}
\end{equation}
Using \cref{eq:memenergy8}, we then compute the non-zero components of the first Piola-Kirchhoff stress $\mathbf{P} = \partial W/\partial \mathbf F$ in the membrane as
\begin{equation}
P_{i} = \pdv{W}{\lambda_{i}} = E \frac{\Lambda^{-1} \left(\lambda_{i}^2 - \Lambda\right)\left(1+\alpha\left(\lambda_{i}^2\Lambda 
 - 1 \right)\right)}{2\lambda_{i}^3\left(1+\nu\right)}, \qquad i=1,2,
\label{eq:1PKmem9a}
\end{equation}
and the components of the principal Cauchy stress $\mathbf{T} = J^{-1}\mathbf{P}\mathbf{F}^{T}$ associated with the membrane energy, as
\begin{equation}
T_{i} = E\frac{\Lambda^{-\frac{3}{2}}\left(\lambda_{i}^2 - \Lambda\right)\left(1 + \alpha\left(\lambda_{i}^2\Lambda - 1\right)\right)}{2\lambda_{i}^2\lambda_1\lambda_{2}\left(1+\nu\right)}, \qquad i=1,2.
\label{eq:cauchystressmem9b}
\end{equation}
Note that in axially symmetric membranes, the Cauchy stress has only in-plane components $(T_1,T_2)$ representing the principal stresses along meridians and parallels, respectively. With stretches given by \cref{eq:prinstrch3}, these are functions of $R$ only.

It is worth noting that the Cauchy stress components \eqref{eq:cauchystressmem9b} satisfy the {\it Baker-Ericksen inequality} $(T_2-T_1)(\lambda_2-\lambda_1)>0$ for all choices of the material parameters. This condition is fundamental to ensure that an inflated, elastically homogeneous spherical membrane in its reference configuration remains spherical for all values of applied pressure, see \cite{DPZse} and \cite{DeTommasi2013}. 
In this paper we demonstrate that an initially flat membrane with elastic inhomogeneities can attain non-trivial geometries (other than the spherical configuration) under a uniform pressure load. Furthermore, wrinkles and necks may be achieved at desired locations.

\subsection{Tension Field Theory: Relaxed strain energy functional}

Ideally, membranes with negligible bending stiffness may only achieve non-negative stress states as they cannot sustain in-plane compression. Real membranes, however, possess a small (albeit non-negligible) bending stiffness which effectively slightly delays the onset of wrinkling when compressive stresses arise. 

Lack of resistance to compression may be seen as a unilateral constitutive constraint. This is embedded in tension field theory by constructing a ``relaxed strain energy density'' $W^{\star}(\lambda_1,\lambda_2)$ from the parent energy $W(\lambda_1,\lambda_2)$, which sets an in-plane stress component to zero whenever it would be negative in the parent energy. Clearly, in taut regions $W^{\star} \equiv W(\lambda_1,\lambda_2)$, whereas in completely slack regions $W^{\star}=0$. 

Following A.C. Pipkin \citep{Pipkin1986TheRE}, we may formalize the above by taking
\begin{equation}
W^{\star}=\begin{cases}
\begin{array}{lllll}
W\left(\lambda_{1}, \lambda_{2}\right) & \text { if } & \lambda_{1} \geq \lambda_{1}^{\star}\left(\lambda_{2},\nu\right), & \lambda_{2} \geq \lambda_{2}^{\star}\left(\lambda_{1},\nu\right), & \textbf{Taut region with biaxial tension}, \\

W\left(\lambda_{1}, \lambda_{2}^{\star}\left(\lambda_{1}, \nu \right) \right) & \text { if } &  \lambda_{1} \geq 1, &   \lambda_{2}  \leq \lambda_{2}^{\star}\left(\lambda_{1}, \nu\right), & \textbf{Wrinkled region with uniaxial tension},\\

W\left(\lambda_{1}^{\star}\left(\lambda_{2}, \nu\right), \lambda_{2}\right)  &  \text { if } &  \lambda_{2} \geq 1, &  \lambda_{1} \leq \lambda_{1}^{\star}\left(\lambda_{2}, \nu\right), & \textbf{Wrinkled region with uniaxial tension}, \\

0  &  \text { if } &  \lambda_{1} \leq 1, &   \lambda_{2}\leq 1, & \textbf{Slack region with no tension}.
\end{array}
\end{cases}
\label{eq:relaxenergy10}
\end{equation}

The function $\lambda_i^{\star}(\lambda_j,\nu)$ is called ``natural width in tension'' and is the main player of the relaxed energy construction. Now consider a strip of membrane with energy \Cref{eq:memenergy8}, oriented along the principal directions. If the membrane is pulled by so that $\lambda_1\geq 1$ along the direction $1$, while being free in the direction $2$, the membrane contracts laterally with stretch ratio $\lambda_2=\lambda_2^{\star}(\lambda_1,\nu)$. This value is found by solving $T_{2}(\lambda_1,\lambda_2)$ = 0 for $\lambda_2$. Therefore if $\lambda_2\leq \lambda_2^{\star}(\lambda_1,\nu)$, then $T_{2}\leq 0$, and the membrane would be compressed in the non-relaxed energy. 
In tension field theory, this problem is avoided by assuming that if the membrane is compressed further than $\lambda_2^{\star}$, its energy does not change once $\lambda_2^*$ is attained: namely, $W^{\star}(\lambda_1,\lambda_2)=W(\lambda_1,\lambda_2^{\star}(\lambda_1,\nu))$. 

Physically, if $\lambda_1$ is kept fixed, there would be no energetic expenditure in shortening the membrane in the direction $2$, below the natural width $\lambda_2^{\star}$. The same considerations apply to the perpendicular direction, whereas in the fully slack region, the energy is directly set to zero. For auxetic materials ($\nu<0$), ``lateral contraction'' is changed to ``lateral expansion'', while all the remaining considerations are unchanged.

Remarkably, for membranes with energy defined by \Cref{eq:memenergy8}, the natural width depends on the Poisson ratio $\nu$ only, as follows, 
\begin{equation}
\begin{split}
    \lambda_{i}^{\star}\left(\lambda_{j}, \nu\right) = \left(\lambda_{j}\right)^{-\nu}, \qquad i,j = 1, 2, \qquad i\neq j.
\end{split}
\end{equation}
Note that for $\nu=\nicefrac{1}{2}$, we recover the expression for the natural width obtained in \cite{steigmann1989finite} and \cite{Steigmann1989wrinkling} for incompressible and isotropic membranes. 

The dependence of the natural width in tension on the Poisson coefficient $\nu$ has not been explored much so far. However, this feature can be powerfully exploited to achieve non-trivial wrinkling patterns on demand. Indeed, as already illustrated by \citet{sai2023} for stretched membranes,  
by carefully tuning the spatial distribution of $E$ and $\nu$, one can achieve unusual wrinkling patterns. This has a great potential in technological applications \citep{naebe2016functionally, ren2018auxetic}.

 Auxetic materials have interesting properties: when $\lambda_{1}>1$, then $\lambda_{2}^{\star}>1$, indicating that the membrane expands in all directions. This is a very strong difference between auxetic and classical membranes. Classical membranes always contract laterally when pulled in the perpendicular direction. In the next section, we discuss the implications of these features for  both classical and auxetic functionally graded membranes.


\subsection{Equilibrium of a pressurized membrane\label{equil-eqns}}


Equilibrium equations of axially symmetric pressurized membranes are written along the meridian and normal directions \citep{green1960large,gurtin1975continuum, libaisimmonds1998}. If the reference membrane is flat, all involved fields depend on the reference radius $R$ only, and the equilibrium equations take the form
\begin{equation}
T_1' + \frac{\rho'}{\rho} \left(T_{1} - T_{2}\right)= 0,
\qquad
\kappa_{1}T_{1} + \kappa_{2}T_{2} = -P,
\label{eq:cireq12d}
\end{equation}
where $\left(\bullet\right)'=d/dR$ and where the membrane curvatures $\kappa_1$ (curvature along a meridian line) and $\kappa_2$ (curvature along a parallel line) in axial symmetry may be calculated as
\begin{equation}
\begin{split}
\kappa_{1} = \frac{\rho'\eta'' - \rho'' \eta'}{\lambda_1^3} = \frac{(\lambda_{2}R)^{'}\lambda_{1}^{'} - (\lambda_{2}R)^{''}\lambda_{1}}{\lambda_{1}^2\sqrt{\lambda_{1}^2 - \big((\lambda_{2}R)^{'}\big)^2}}, \qquad
\kappa_{2} = \frac{\eta'}{\rho \lambda_1}
 = -\frac{\mathbf{n}\cdot\mathbf{e}_{r}}{\rho} =  -\frac{\sqrt{\lambda_{1}^2 - \big((\lambda_{2}R)^{'}\big)^2}}{\lambda_{1}\lambda_{2}R}.
\end{split}
\label{eq:curvprin15}
\end{equation}
Note that for $\kappa_2$ we have also used the alternative expression based on the outward normal $\mathbf{n}$ to the current surface, and on the radial unit vector $\mathbf{e}_r$ pointing outwards radially and perpendicularly to the membrane axis $z$. This expression exemplifies that the curvature of a parallel {\it does not coincide}, in general, with the curvature of the membrane along the parallel unless of course $\mathbf{n}\equiv\mathbf{e}_r$. 

Because the membrane is flat in its reference configuration, the fields $\rho,\eta$ must satisfy the following boundary conditions,
\begin{equation}\label{bcs}
\rho(R_\text{fin}) = 0, \qquad
\eta(R_\text{fin}) = 0. 
\end{equation}
Also, because point loads are not applied to the membrane, the suitable boundary conditions to be imposed on the functions $\rho,\eta$ to avoid stretch and curvature singularities at the origin are
\begin{equation}\label{conds}
\lambda_1(0)=\lambda_2(0),\qquad
\kappa_1(0)=\kappa_2(0). 
\end{equation}

Finally, to account for the presence of {\it wrinkling}, it is sufficient to write the equilibrium equations in terms of the relaxed counterparts $(T_1^{\star},T_2^{\star})$ as obtained from the relaxed energy through $\mathbf{T}^{\star} = J^{-1}(\partial W^\star/\partial \mathbf F)\mathbf F^T$. This will automatically ensure that no compressive states can be achieved on the inflated membrane. 

\section{General insights into necks and wrinkles\label{insights}}


{\color{black} Due to constitutive and geometric non-linearities, it would appear that little can be said in general  on the placement of wrinkles and necks in an inflated membrane undergoing large deformations. Surprisingly, a careful analysis of the equilibrium equations of an inflated membrane, together with the implementation of tension field theory, provides a deep and fully general characterization of what type of instability patterns may or may not be expected in an inflated membrane. 

Even more remarkably, these characterizations are {\it universal}, in the sense that they are independent on the choice of the (isotropic and elastic) constitutive behavior of the material. One such remarkable universal result is that {\it necks and wrinkles can never overlap}: this insight will be used in the sequel to produce alternating patterns of regions with wrinkling and necking. 

In this section we collect these universal characterizations, calling them ``{\it insights}'', that apply to  {\it all isotropic, axisymmetric and inflated membranes.} 
Their relevance to the present study is that one can use them as general guidelines to design the spatial distribution of elastic moduli, in order to obtain desired patterns of wrinkles and necks in the inflated membrane. 
The insights also provide interesting characterizations of the membrane shape when wrinkles or necks occur, therefore helping to understand what type of shapes may, or may not, be obtained in an inflated membrane undergoing such instabilities. }

As we deal with thin membranes that offer no resistance to compression, we use tension field theory and base our analysis on the equilibrium equations \cref{eq:cireq12d}, expressed in terms of the relaxed Cauchy principal stress components $T_1^{\star},T_2^{\star}$. 

\begin{itemize}
\item{\textbf{Insight 1:}} {\it The membrane curvature along the parallels is always negative: $\kappa_2<0$}.

\medskip

First note that the equilibrium of a cap above a parallel can be written, in global form, by balancing pressure $P>0$ and membrane tension, to give
\begin{equation}
T_{1}^{\star}\kappa_{2} + \frac{P}{2} = 0. 
\label{eq:stress116b}
\end{equation}
Because, by definition, $T_{1}^{\star}\geq 0$, equilibrium imposes $T_{1}^{\star}>0$ and $\kappa_2<0$ everywhere, which concludes the proof. 

\medskip

In passing, note that $\kappa_{2} = \eta'/(\rho \lambda_1)$ so that $\eta(R)$ is a decreasing function from $R=0$ to $R=R_\text{fin}$. But $\eta(R_\text{fin})=0$, which means that $\eta(0)>\eta(R)>0$, in agreement with the fact that a membrane inflated from below ($z<0$) displaces upwards. 

\medskip

\item{\textbf{Insight 2:}}  {\it  Wrinkles can never be aligned with parallels}. 

\medskip

If such wrinkles were to occur it would result in $T_{1}^{\star}=0$, but from Insight 1 we know this is not possible. Therefore, wrinkles can only be aligned with meridians.

\medskip

\item{\textbf{Insight 3:}}  {\it The following inequality is always satisfied in the membrane:}
$\kappa_1\geq - 2|\kappa_2|$.

To prove this, we use \cref{eq:cireq12d} and \cref{eq:stress116b} to get\footnote{Note sign difference with respect to \cite{hill1950c} and \cite{machado2012membrane} due to different notations} 
\begin{equation}
T_{2}^{\star} = -\frac{P}{\kappa_{2}}\left(1 - \frac{\kappa_{1}}{2\kappa_{2}}\right).
\label{eq:stress215}
\end{equation}
Because $T_{2}^{\star} \geq 0$ and $\kappa_2<0$, and by setting $\kappa_2=-|\kappa_2|$,  we obtain 
\begin{equation}
\kappa_{1} \geq - 2|\kappa_{2}|. 
\label{eq:wrinklcurv17}
\end{equation}

It can also be interpreted as ${\kappa_1}/{\kappa_2} \leq 2$, always.

\medskip

\item{\textbf{Insight 4:}}  {\it When wrinkles occur, the membrane curvatures are linked through:} $\kappa_1 = - 2|\kappa_2|$. 

\medskip

This result follows from the previous point with $T_{2}^{\star} = 0$.  

\medskip

\item{\textbf{Insight 5:}} {\it Neck regions (where $\kappa_G=\kappa_1\kappa_2<0$) and wrinkled regions can never overlap.} 

\medskip

If there is wrinkling, then $\kappa_{1}=- 2|\kappa_{2}|<0$; therefore the Gaussian curvature becomes positive, $\kappa_G= \kappa_{1}\kappa_{2} > 0$, and thus, no necking can occur in the region of wrinkling. 

\medskip

\item{\textbf{Insight 6:}}  {\it For necking to happen, the principal stresses must satisfy:}
$T^{\star}_2 \geq  2 T_1^{\star}$.

\medskip

By using \Crefrange{eq:stress116b}{eq:stress215}, we obtain the principal curvatures in terms of stresses
\begin{equation}
 \kappa_{1} = -\frac{P}{T_{1}^{\star}} \left( 1 - \frac{T_{2}^{\star}}{2T_{1}^{\star}}\right), \qquad \kappa_{2} = -\frac{P}{2T_{1}^{\star}}, \qquad \kappa_{G} = \kappa_{1}\kappa_{2} = \frac{P^2}{2{T_{1}^{\star}}^2}\left( 1 - \frac{T_{2}^{\star}}{2T_{1}^{\star}}\right).
\label{eq:necking18}
\end{equation}
For necking we need $\kappa_{G}<0$, and the result follows\footnote{Using this condition, we can design a membrane material with alternating regions of low and high Young's modulus to create alternating regions of necking and wrinkling under a uniform pressure load.}.

This insight may be also used to prove that wrinkles can never overlap necks. Indeed, for necks $2 T_1^{\star} \leq T^{\star}_2$, for wrinkles $T^{\star}_{2}= 0$, and combining both we get $2 T_1^{\star} \leq T^{\star}_2=0$, which cannot occur because $T_1^{\star}$ must be strictly positive.

\medskip

\item{\textbf{Insight 7:}}  {\it Neither wrinkling nor necking can take place at the apex of the membrane.} 

\medskip

\noindent According to \cref{conds}, both  principal stretches are equal at the center of the membrane, i.e.  $\lambda_1=\lambda_2$ at $R=0$. Therefore, we have $T_1^{\star}=T_2^{\star}$ and $\kappa_1=\kappa_2$ at $R=0$. Then, from Insights 4 and 6, we conclude that neither wrinkling nor necking can occur at the apex of the membrane.

\end{itemize}

Based on Insight 2, we therefore conclude that slack regions and regions of wrinkles aligned with parallels can never occur in the inflated membranes. These observations help us to simplify the relaxed energy function in \cref{eq:relaxenergy10} as follows,
\begin{equation}
W^{\star}=\begin{cases}
\begin{array}{lllll}
W\left(\lambda_{1}, \lambda_{2}\right) & \text { if } & \lambda_{1} \geq \lambda_{1}^{*}\left(\lambda_{2},\nu\right), & \lambda_{2} \geq \lambda_{2}^{*}\left(\lambda_{1},\nu\right), & \\
W\left(\lambda_{1}, \lambda_{2}^{*}\left(\lambda_{1}, \nu \right) \right) & \text { if } &  \lambda_{1} \geq 1, &   \lambda_{2}  \leq \lambda_{2}^{*}\left(\lambda_{1}, \nu\right).
\end{array}
\end{cases}
\label{eq:relaxenergy10b}
\end{equation}

We emphasize here that not all insights are directly helpful in providing  guidance on how to design the functionally graded membrane to get desired instabilities. From Insights 1, 2, 3, 4, 5, 7 we deduce important information on the shape of an inflated membrane, with remarkable bounds on the membrane curvatures in regions with and without instabilities. These insights dictate what shapes may be expected (or not expected) in an inflated membrane, regardless of its constitutive response. 

From Insight 6, together with the consideration that $T_2^{\star}=0$ leads to the appearance of wrinkles, we conclude that it is possible to create necks and wrinkles by tuning the ratio between $T_2^{\star}$ and $T_1^{\star}$. This method is used in the following sections to create instability patterns at desired locations. 




\section{Designing instabilities\label{results1}}



In principle, it is logical to assume that the spatial distribution of the elastic moduli $E(R), \nu(R)$ may be used to achieve wrinkles and necks. However, in practice, the inverse problem is extremely difficult due to geometric and constitutive nonlinearities. To circumvent the analytical complexity of this inverse problem, we use numerical simulations to devise regions of spatial inhomogeneity with sharp variations of the elastic moduli to get desired instabilities.

Across a boundary of radius $R$ between elastically inhomogeneous regions, the following jump conditions have to be satisfied, 
\begin{equation}
[T_1^{\star}]= 0, \qquad [\lambda_2] = 0 
\end{equation}
where $[f]:=\lim_{\epsilon\rightarrow 0} (f(R+\epsilon)-f(R-\epsilon))$. We see that across such boundary, we may sharply act on the jump of the tension $[T_2^{\star}]$ in the direction of parallels, while at the same time, the value of $T_1^{\star}$ along meridians shall remain continuous across the phase boundary. By tuning the ratio ${E_{\text{stiff}}}/{E_{\text{soft}}}$ between the Young's modulus of two neighboring regions (stiff and soft), we can act on the ratio $T_2^{\star}/T_1^{\star}$ which, as the insights above reveal, regulates the onset of wrinkles (when $T_2^{\star}/T_1^{\star}$ is decreased toward 0) or the onset of necks (when $T_2^{\star}/T_1^{\star}$ is increased above 2). 

The Poisson coefficient $\nu$ is also a key player. Indeed, by taking this coefficient to be negative, the membrane will expand laterally and create compressive stresses  ($T_2^{\star}=0$). 

The complexity of the problem requires a trial and error approach to determine appropriate elastic inhomogeneity patterns to achieve the sought instabilities. By using the commercial software \texttt{COMSOL}, we implement the Blatz-Ko relaxed strain energy function, \cref{eq:relaxenergy10}, for functionally-graded, inflated membranes.

In the simulations below, we inflate  pre-stretched membranes by keeping the external rim fixed. 
Then we solve \cref{eq:cireq12d} to obtain the Pressure-Volume (P-V) curves and the deformation profiles of the membranes. The numerical implementation is explained in \cref{numericalmodel}.
Our simulations are consistent with the universal insights developed above: notably, we see that necking cannot occur in wrinkling regions and that only wrinkles aligned with meridians can occur. 
By playing on the ratio ${E_{\text{stiff}}}/{E_{\text{soft}}}$ and on the location of elastically inhomogeneous regions, we show that  a desired number of wrinkling and necking regions can be obtained at prescribed locations.

For the sake of brevity, the effects of pre-stretch and of the Blatz-Ko material parameters are not discussed in detail. Here, we focus on the most relevant design parameters: $\nu$ and ${E_{\text{stiff}}}/{E_{\text{soft}}}$.

\bigskip



\subsection{Necks and no wrinkles\label{necks}}

\begin{figure*}[thb]
\centering
\includegraphics[scale=0.5]{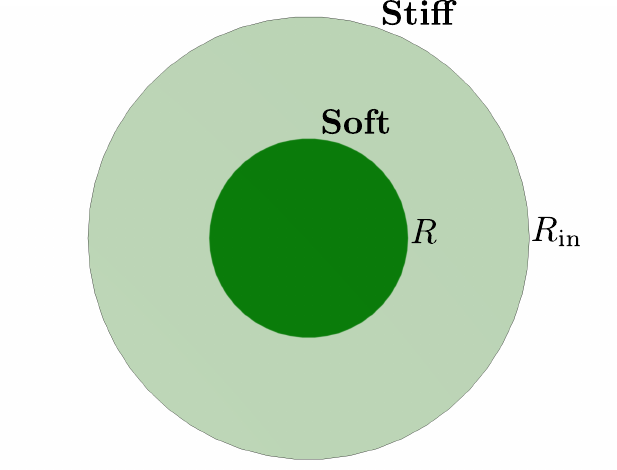}
    \caption{Schematic representation of the one-step variation of the Young modulus distribution $E$ in the reference configuration. The circular membrane has a softer region in the center surrounded by a stiffer region that extends to the disc circumference, see \cref{eq:appenE1SV22a}.}
    \label{fig:YM-neck}
\end{figure*}

To ensure that the inflated membrane develops necks without wrinkles, we have to create regions where the ratio $T_2^{\star}/T_1^{\star}$ exceeds 2, while maintaining $T_2^{\star}>0$ everywhere in the membrane. To achieve this goal, in \cref{necks}, we use a one-step variation of the Young modulus. The reference membrane has a central region with low Young's Modulus ($E_{\text{soft}}$) surrounded by a region of high Young's Modulus ($E_{\text{stiff}}$) which extends up to the boundary, as shown in \cref{fig:YM-neck}. Moreover, to avoid that $T_2^{\star}$ approaches zero, we set $0< \nu < -0.6$ which ensures a low-to-moderate auxeticity.
 
\begin{figure*}[thb]
\centering
\includegraphics[scale=0.95]{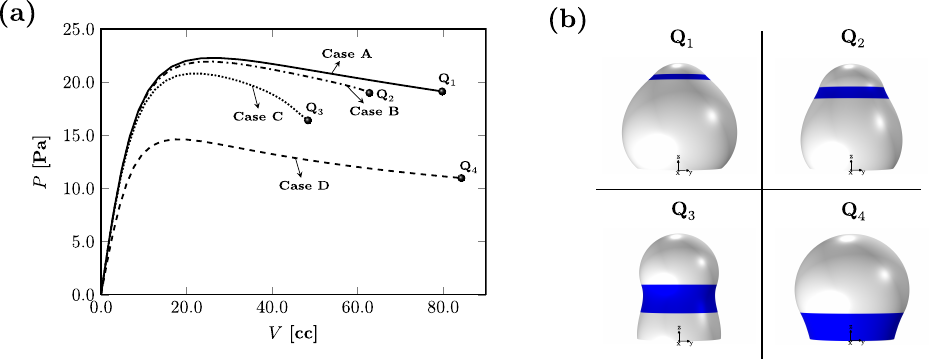}
    \caption{(a) Inflation process: Pressure-Volume curves of a pre-stretched circular membranes with four different cases of one-step variation of the Young modulus. From Case A 
 to Case D, the radius or size of $E_{\text{soft}}$ region is assumed to be increasing (see \cref{eq:appenE1SV22b}),  (b) Deformation profiles of the membrane at the end of inflation process are shown here for each case. Inflation process of each membrane is continued until the apex of the membrane reaches a height of five times the value of radius $R_{\text{in}}$. For each profile, the region in blue highlights the necking area ($\kappa_{\pazocal{G}} <0$). Blatz-Ko parameter $\alpha = 0.4$, pre-stretch $\lambda_{p}=2$, Poisson's ratio $\nu = -0.4$, and referential radius $R_{\text{in}}= 1 \hspace{1mm} \text{cm}$.}
    \label{fig:Only-necks}
\end{figure*}

Our simulations show that if $E_{\text{stiff}}/E_{\text{soft}}$ is small (< 1.5), the membrane displays a smooth necking during inflation. If $E_{\text{stiff}}/E_{\text{soft}}$ is moderate ($1.5 < E_{\text{stiff}}/E_{\text{soft}} <3$), the necking region is more prominent.

On the left panel of \cref{fig:Only-necks}, we plot the P-V profiles of pre-stretched circular membranes with different sizes of the $E_{\text{soft}}$ region. From Case A 
 to Case D, the radius or size of $E_{\text{soft}}$ region increases (see \cref{eq:appenE1SV22b}). We observe that, as the size of the soft region increases, the value of the limit-point pressure $P_{\text{lim}}$ decreases. This clearly occurs because the bigger the soft region, the smaller the effective Young modulus.

 The deformation profile at the end of inflation process is shown for each case on the right panel of \cref{fig:Only-necks}. It can be clearly noticed that, by adjusting the size of the low Young modulus region, we can shift the boundary of the necking region. Therefore necks may be obtained at any desired location. It has to be noted that the simulations confirm that necking cannot occur at the tip of the membrane, in agreement with Insight 7.
\subsubsection*{Special case: Double bubbling}

\begin{figure*}[thb]
\centering
\includegraphics[scale=0.75]{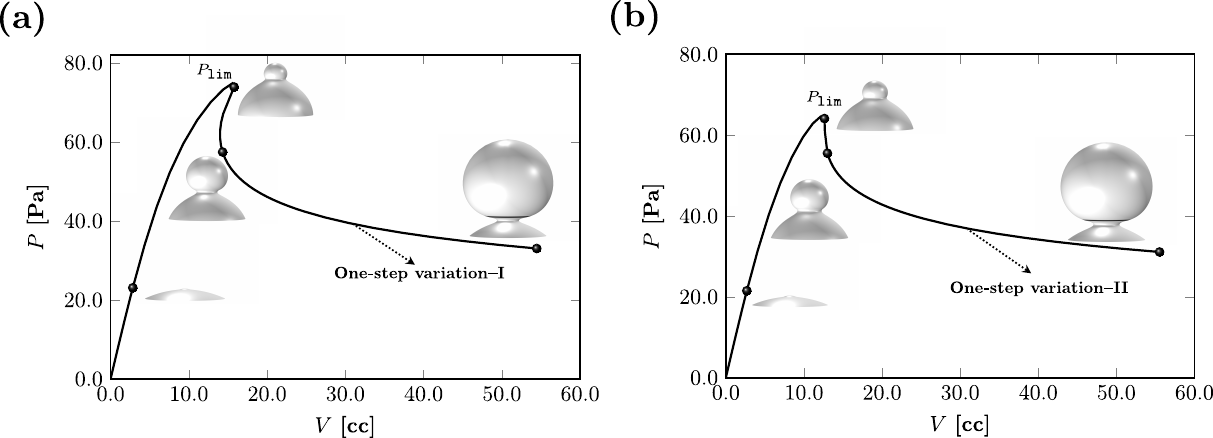}
    \caption{Pressure-Volume curves of pre-stretched circular membranes for two different cases of one-step variation of the Young modulus (see, \cref{eq:appenE1SV22c}) are shown: (a) One-step variation--I, (b) One-step variation--II. The deformation profiles of the membrane are shown at four different points along the inflation path. The black region in the deformation profiles (at the end of inflation process) denotes wrinkling regions. Parameters: $R_{\text{in}}= 1 \hspace{1mm} \text{cm}, \lambda_p=2$, $\alpha = 0.4$ and $\nu = -0.9$.}
    \label{fig:double-bubbles}
\end{figure*}

Membranes with a high stiffness ratio (${E_{\text{stiff}}}/{E_{\text{soft}}}$ >3) undergo a transition into a {\it double-bubbled shape}, due to a sudden inflation of the softer region after the limit-point pressure, as shown in \cref{fig:double-bubbles}. In this figure, we plot the P-V curves for two different cases of one-step variation of the Young modulus.

It is important to note that the behavior of the P-V curve is strongly dependent on the size of the softer core. In \cref{fig:double-bubbles}(a), the volume decreases immediately after the limit point, whereas in \cref{fig:double-bubbles}(b) the volume keeps increasing while the pressure decreases after the limit point. Note that case (a) is also studied by \cite{selvadurai2022mechanics} for pressurized incompressible elliptical membranes with the Ogden strain energy model.

We also show the deformed shapes at four different stages of inflation. Note that the formation of the second bubble corresponds to the necking point on the P-V curve.

Highly auxetic membranes (with $\nu<-0.8$) develop wrinkles in the region between the two bubbles. Indeed,  high auxeticity results into high compressive stresses.  Note that in line with Insight 5, wrinkles (black color regions) are never observed in the necking region, but only slightly above the neck.

\subsection{Necks and wrinkles} \label{circase3} 

In this section, we extend the findings from \cref{necks} to the general case where the Young modulus has an $n$-step variation across the membrane, while $\nu=-0.9$ is uniform throughout the membrane. Again, we use alternating stiff and soft regions to create multiple necks. Depending on the size of these regions we may or may not achieve lateral compression, and therefore wrinkling, in the membrane. Interestingly, the complexity of the problem leads to wrinkling regions appearing and disappearing during the inflation process.

\begin{figure*}[!ht]
\centering
\includegraphics[width=0.52\textwidth]{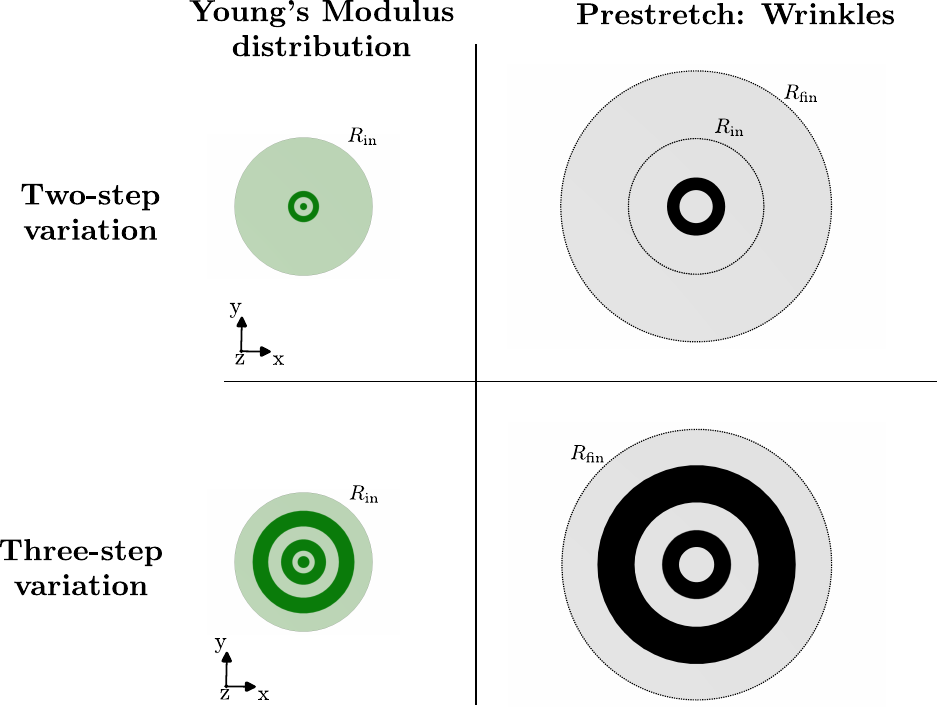}
    \caption{First column: two-step (\Crefrange{eq:appenE2SV23a}{eq:appenE2SV23b}) and three-step (\Crefrange{eq:appenE3SV24a}{eq:appenE3SV24b}) variations in the Young modulus $E$. 
    Right column: the corresponding wrinkling profiles appearing during the pre-stretch process. 
    Here, the pre-stretch is $\lambda_p=2$, with material parameters: $\alpha=-0.4$ and $\nu=-0.9$.}
    \label{fig:EvsR-2-3-SV}
\end{figure*}

\begin{figure*}[!ht]
\centering
\includegraphics[scale=0.75]{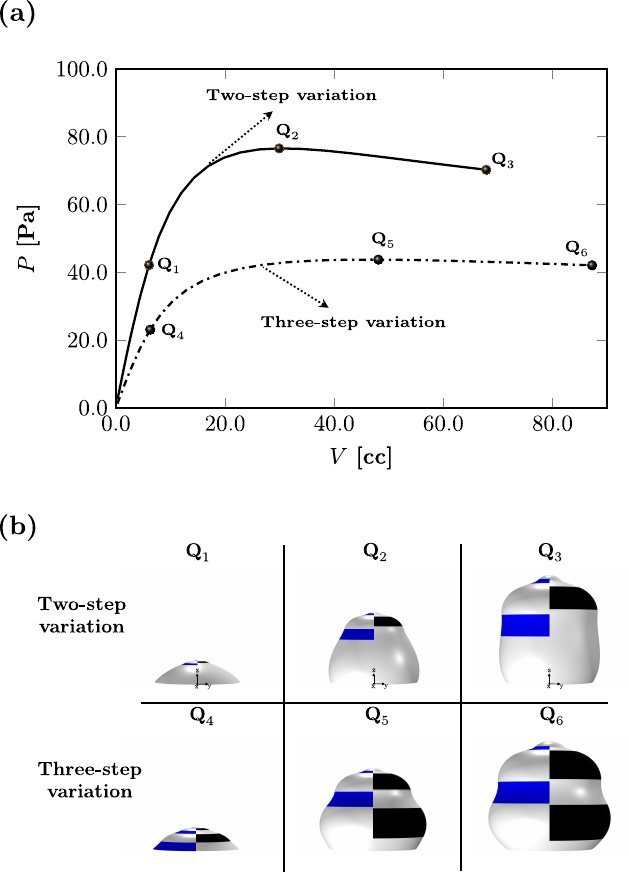}
   \caption{(a) Inflation process: Pressure-Volume curves of a pre-stretched circular membrane with two-step (solid curve) and three-step (dash-dotted curve) variations of the Young modulus, (b) Deformation profiles of the membrane at three different locations are shown for each variation case. For each profile, the left half highlights the necking area in blue ($\kappa_{\pazocal{G}} <0$) and the right half highlights the wrinkling area in black ($T_{2} = 0$).  
   Parameters: $\lambda_p=2$, $\alpha = 0.4$ and $\nu = -0.9$.}
    \label{fig:PvsV-2-3-SV}
\end{figure*}


The simulations show that membranes with $n$-step variations of the Young modulus develop $n-1$ wrinkled regions  during the pre-stretching process, consistent with the findings of \cite{sai2023}. We recall that during pre-stretching, the membrane remains flat. These simulations are reported in \cref{fig:EvsR-2-3-SV}, showing the onset of wrinkling patterns in laterally pre-stretched membranes with two- and three-step variations of the elastic moduli.

Following the pre-stretch, we simulate the inflation. The deformation profiles for the inflation process are shown in \cref{fig:PvsV-2-3-SV}(b). Wrinkling and necking regions are represented by black (right half of the figure) and blue (left half) colors, respectively. Our simulations show that stiff regions inflate relatively less than soft regions. Moreover, this difference leads to the formation of a neck at the boundary between the two regions. Interestingly, the wrinkles generated during the pre-stretch phase tend to be preserved during the inflation process. Also, the wrinkles never overlap with the necking regions, which is consistent with the findings of Insight 5.

Furthermore, depending on the stiffness ratio between soft and stiff regions, the spacing between the necks and the wrinkles can be modulated. In particular, the higher the ratio ${E_{\text{stiff}}}/{E_{\text{soft}}}$, the closer the two patterns are, and vice-versa.

The spatial heterogeneity of the elastic moduli affects also the P-V curve, as shown in \cref{fig:PvsV-2-3-SV}(a). Interestingly, for membranes with two-step variation of the Young modulus, the pressure increases until the limit-point instability and later drops, whereas it remains nearly constant in the three-step variation case. In all simulations, no further increase of pressure past the limit point is observed, irrespective of the material constant $0<\alpha<1$.

Finally, we highlight two important features. 
First, configurations containing multiple regions of necks and wrinkles can be obtained by increasing the number of stiff-soft regions in the original membrane.
Second, the pre-stretch $\lambda_p$ may be used as a parameter to \textit{unwrinkle}, during inflation, the regions of the membrane that are close to the boundary. 


\subsection{Wrinkles and no necks\label{wrinkles}}

\begin{figure*}[thb]
\centering
\includegraphics[scale=0.85]{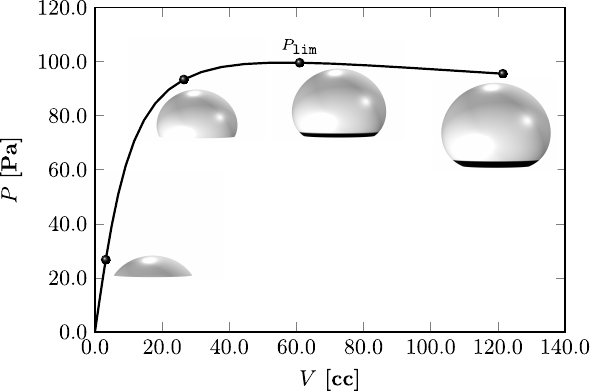}
    \caption{Pressure-Volume curve of pre-stretched circular membranes with constant material properties. The deformation profiles of the membrane are shown at four different points along the inflation path. The black region in the deformation profiles denotes wrinkling regions. Parameters: $R_{\text{in}}= 1 \hspace{1mm} \text{cm}, \lambda_p=2$, $\alpha = 0.4, E=70$ kPa, and $\nu = -0.9$.}
    \label{fig:PVcons}
\end{figure*}

Our results from the previous sections suggest that a high contrast between stiff and soft regions (i.e. high ${E_{\text{stiff}}}/{E_{\text{soft}}}$)  leads to necking. Therefore, to avoid necks, we now consider a homogeneous (or nearly homogeneous) distribution of the Young modulus across the membrane. Although different types of distributions (constant, linear, step, and Gaussian) of the Poisson ratio are explored for this case, they all yield similar deformation behaviors. 

For the sake of brevity, we consider the case where the material properties $E, \nu$ are spatially uniform throughout the membrane. We also compare our results with existing solutions in the literature.

In \cref{fig:PVcons}, we see that for an auxetic membrane with homogeneous Young's modulus, the pressure increases with the volume up to the limit point, where the membrane loses stability and wrinkles appear near the base of the membrane. 
Beyond the limit point, the pressure decreases with increasing volume. The results in \cref{fig:PVcons} also show that the membrane attains a spherical shape, and indeed no necks are developed. These findings are consistent with previous experiments on pressurized incompressible circular membranes \citep{treloar1944strains, machado2012membrane, zhou2018evaluation} and with several analytical and numerical studies \citep{yang1970axisymmetrical,patil2013finite,selvadurai2022mechanics}.

For auxetic membranes, we predict the formation of wrinkles aligned with meridians. 
We note that \citet{chaudhuri2014static} predict wrinkles aligned with parallels in inflated incompressible membranes with homogeneous material properties. However, according to our Insight 2,  such wrinkles are not possible in any axisymmetrically inflated isotropic circular membranes. 

Additionally, we find that for a Blatz-Ko material with homogeneous material properties, tensile stresses exist everywhere in the membrane, except near the fixed circumferential edge. Hence, wrinkles are obtained only near the fixed boundary. 

Finally, we mention that wrinkles, if present, may also be suppressed by tuning the Poisson ratio of the membrane and the pre-stretch applied to the membrane before inflating. We also observed separately that the limit-point pressure $P_{\texttt{lim}}$ increases linearly with the Blatz-Ko material constant $\alpha$ and nonlinearly with the pre-stretch, respectively.

\section{Conclusion and perspectives\label{conclusions}}

\begin{figure*}[!ht]
\centering
\includegraphics[scale=0.8]{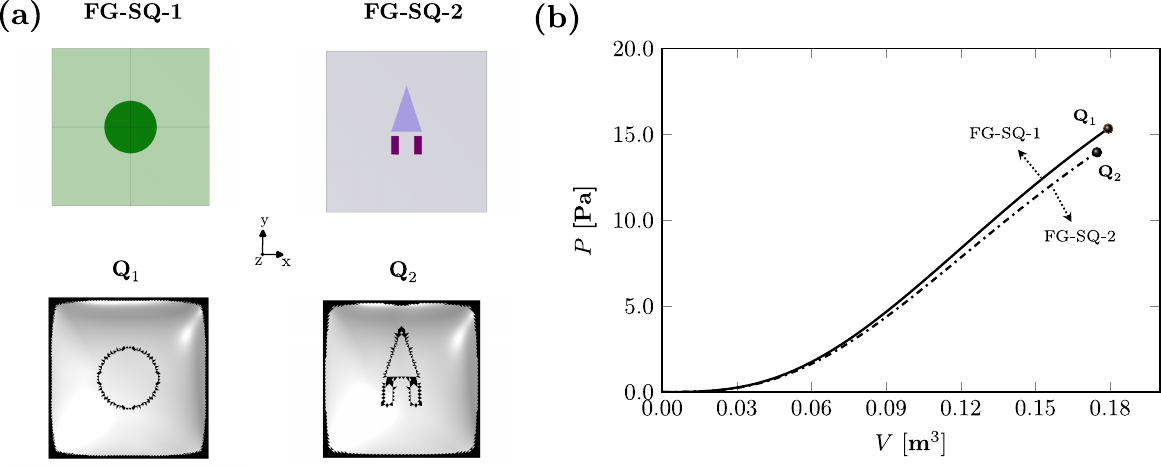}
    \caption{(a) Square membranes with two different Young's modulus profiles (top) and wrinkling patterns (bottom); (b) Associated Pressure-Volume curves. FG-SQ-1: The circular region in the center and the rest of the membrane have $E= 0.3$ MPa and $E=0.03$ MPa, respectively. FG-SQ-2: The triangular and rectangular strips have $E= 0.63$ MPa and $E=4.03$ MPa, respectively, while the rest of the membrane has $E=0.03$ MPa. The Poisson ratio is $\nu=-0.1$, pre-stretch $\lambda_{p}=1$, and the Blatz-Ko coefficient is $\alpha = 0.4$. Wrinkling patterns at two points $\text{Q}_{1}$ and $\text{Q}_{2}$ on the inflation curves corresponding to FG-SQ-1 and FG-SQ-2, respectively, are shown at the bottom of (a).}
    \label{fig:square-EvsX-Y}
\end{figure*}
 
 With this work we investigated the possibility of inducing various forms of instabilities in inflated hyperelastic membranes. Our results are valid for axially symmetric configurations. In particular, we investigated the effects of spatial material inhomogeneities on membranes with auxetic properties ($\nu<0$). 
 
We showed that wrinkling and necking instabilities can be induced in the membrane during inflation, by tuning the material inhomogeneities. We avoided solving  the complex inverse problem of identifying the elastic heterogeneity patterns that deliver the desired instabilities at desired locations. Instead, we developed the  following novel methods:
\begin{enumerate}
\item We implemented tension-field theory based on the relaxed strain energy function into \texttt{COMSOL} to study the stability of an inflated compressible, hyperelastic membrane of the Blatz-Ko type for an arbitrary geometry; 
\item We obtained a series of universal results for the formation of wrinkles and necks in thin inflated membranes; 
\item Building upon the universal results and the numerical simulations, we identified spatial inhomogeneity distributions across the undeformed membrane that result in wrinkles alone, necks alone, and simultaneous (but not overlapping) wrinkles and necks in the inflated membrane. 
\end{enumerate}

Our study is limited by various factors. For example, typical auxetic membranes are anisotropic in nature and their material properties are deformation-dependent: these features might greatly affect the results presented in this work. This aspect is not addressed in the current work but could be studied with the methods we have developed here. 

Another avenue of interest is to investigate the inflation of non-axisymmetric membranes. For example, in \cref{fig:square-EvsX-Y}, we show  2D distributions of material properties in square membranes. 
The membrane is fixed on its edges and inflated under a uniform pressure load. We show that material properties can be tuned in the membranes to produce desired wrinkling patterns, as shown in the \cref{fig:square-EvsX-Y}. This concept could potentially be used in Braille reading and haptics.

\printcredits

\section*{Declaration of Competing Interest}
The authors declare that they have no known competing financial interests or personal relationships that could have appeared to influence the work reported in this paper.

\section*{Acknowledgments}
\begin{figure*}[!ht]
\centering
\includegraphics[width=0.15\textwidth]{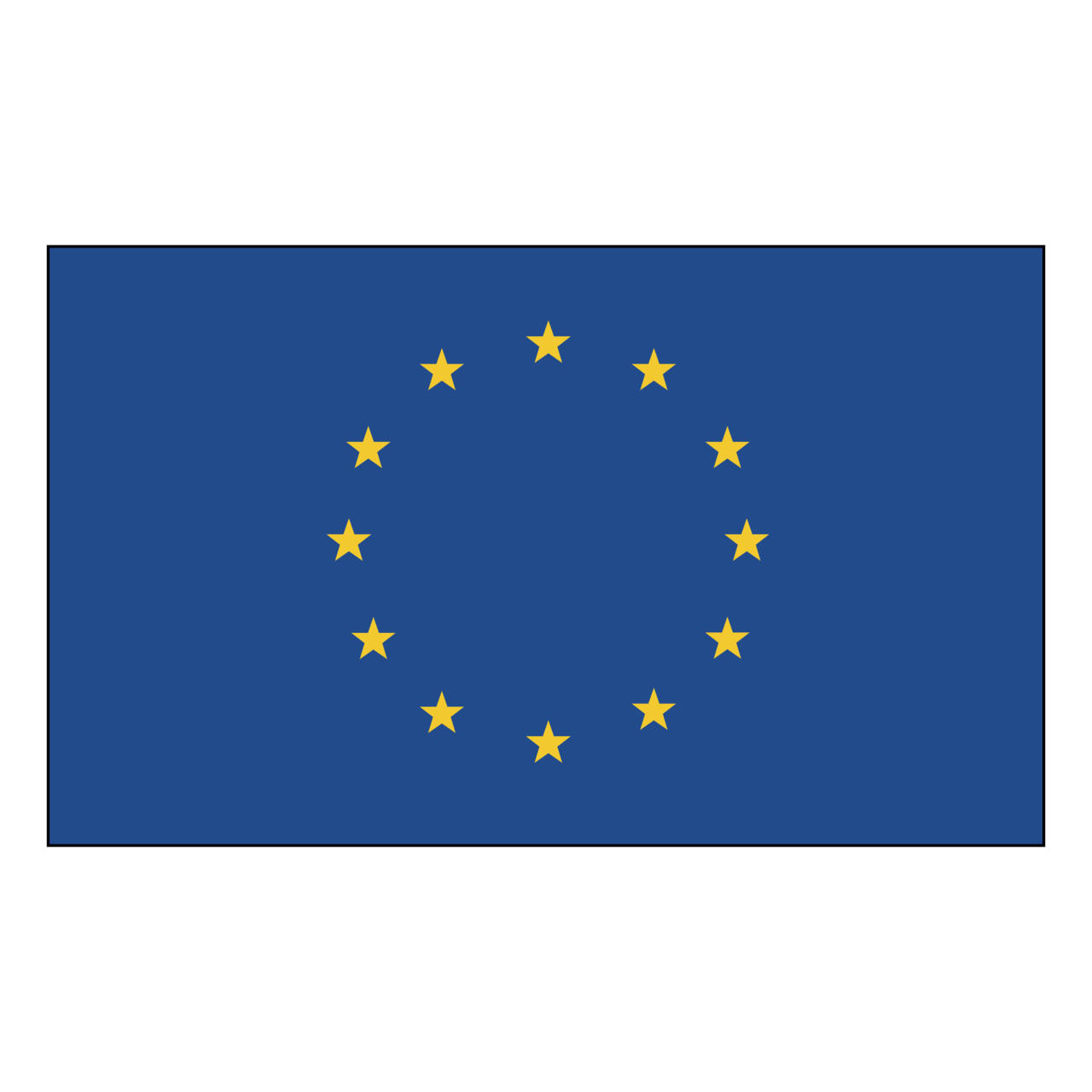}
\end{figure*}
This project has received funding from the European Union’s Horizon 2020 research and innovation programme under the Marie Skłodowska-Curie Grant Agreement No. 956401. GZ gratefully acknowledges the support of GNFM (Gruppo Nazionale di Fisica Matematica) of the INdAM F. Severi. 
\section*{Supplementary data}
No supplementary material is attached to this article.


\appendix
\section{Appendix: Numerical model in \texttt{COMSOL}}\label{numericalmodel}
\setcounter{equation}{0}
\renewcommand{\theequation}{A.\arabic{equation}}

We adjust the pre-stretch $\lambda_{p}$ by changing the radius of the undeformed membrane $R_{\text{in}}$, while holding the radius of the pre-stretched membrane $R_{\text{fin}}$ fixed. 
Following \cite{gonccalves2009nonlinear, patil2013finite}, we represent the radial position of point $\pazocal{P}_{2}$ in the stretched membrane as
\begin{equation}
\begin{split}
 \rho_{0} &= \lambda_{p} R, \qquad \lambda_{p} = \frac{R_{\text{fin}}}{R_{\text{in}}}.      
\end{split}
\label{eq:radprestretch1}
\end{equation}

We assume that the pre-stretched circular membrane is fixed on its boundaries, i.e.
$\mathbf{u}(R_{\text{fin}}) = 0$, where $\mathbf{u}$ is the displacement field in the deformed membrane and $R_{\text{fin}}$ is the radius of the pre-stretched circular membrane.\par

The initial thickness of the circular membrane is considered to be negligible in comparison to its radius (we take $H = R_\text{in}/200$).   
We take the Young modulus to vary spatially as a function of the referential radial coordinate $R$. 

From an experimental viewpoint, the volume-controlled inflation process is performed by increasing the mass of the gas in the membrane \citep{ericksen1998introduction}. However, in our simulations, we do not model the behavior of the gas. 
Instead, we implement the inflation by using a volume-controlled (or displacement-controlled) procedure as proposed by \citet{yang1970axisymmetrical}, \citet{pujara1978deformations}, and \citet{patil2013finite}. We prescribe a vertical displacement at the center of the membrane and calculate the corresponding pressure through a global optimization process using an inbuilt function in \texttt{COMSOL}.\par
The volume enclosed by the surface of the deformed membrane is computed as:
\begin{equation}
V = \frac{1}{3} \int_{\Omega} \left(\pazocal{Q} \cdot \mathbf{n}\right) \text da,
\label{eq:volcal21}
\end{equation}
where $\Omega$ represents the surface domain of the deformed membrane. The position vector of any point on the deformed surface of the membrane is represented by $\pazocal{Q}$ with $\mathbf{n}$ being the outward unit normal at that membrane point.

Here we briefly summarize the main features of that relaxed strain energy implementation in \texttt{COMSOL} software for axisymmetric circular membrane (see \citet{sai2023}):
\begin{enumerate}
    \item By taking advantage of axisymmetry around the Z-axis, we consider a line segment (which lies on mid-plane of the membrane) in $(R, Z)$ plane with membrane elements for our analysis in a cylindrical coordinate system.
    \item The relaxed energy functional (\cref{eq:relaxenergy10}) depends only on the principal stretches and material properties. In \texttt{COMSOL}, by using variable definitions under the component section in the model builder, we define mathematical expressions for the material properties. Principal stretches and invariants of the right Cauchy-Green tensor are then obtained by using internal variables in the software. Once the expressions for natural widths are defined, the relaxed strain energy functional is written in the definition window under the component section, which will be assigned as a user-defined hyperelastic model to the simulation geometry.
    \item Once the boundary conditions and loading are assigned to the geometry with a desired mesh, different numerical solvers can be employed for the analysis.
    \item In the post-processing stage, we obtain the location and orientation of wrinkles using logical operators based on the natural width conditions mentioned in \cref{eq:relaxenergy10}.
\end{enumerate}

\section{Appendix: Circular membranes}\label{appenB}
\setcounter{equation}{0}
\renewcommand{\theequation}{B.\arabic{equation}}

\subsection{One-step variation in material properties} \label{appencircase2} 
The variation in the Young modulus for the membrane is according to the following function:
\begin{subequations}
\begin{equation}
\begin{split}
E(R) &= E_{1}+E_{2}\Bigg(\frac{\exp{A \frac{(R - {Y}_{1})}{D}}-1}{\exp{A \frac{(R - {Y}_{1})}{D}}+1}\Bigg)-E_{2}\Bigg(\frac{\exp{A\frac{(R - {Y}_{2})}{D}}-1}{\exp{A\frac{(R - {Y}_{2})}{D}}+1}\Bigg).
\label{eq:appenE1SV22a}
\end{split}
\end{equation}
\subsubsection{Only necks}
\begin{equation}
\begin{split}
E_{1} &= 10 \hspace{1mm} \text{kPa}, \hspace{2mm} E_{2} = 5 \hspace{1mm} \text{kPa}, \hspace{2mm} {A} = 2, \hspace{2mm} {Y}_{2} = 1.1 \hspace{1mm} \text{cm},  \hspace{2mm} {D} = 0.003 \hspace{1mm} \text{cm}, \\
{Y}_{1} &= 0.07875\hspace{1mm} \text{cm}, \qquad (\text{for Case A}), \qquad {Y}_{1} = 0.105 \hspace{1mm} \text{cm}, \qquad  (\text{for Case B}), \\
{Y}_{1} &= 0.16125 \hspace{1mm} \text{cm}, \qquad (\text{for Case C}), \qquad {Y}_{1} = 0.45 \hspace{1mm} \text{cm}, \qquad  (\text{for Case D}),
\label{eq:appenE1SV22b}
\end{split}
\end{equation}
\subsubsection{Double bubbling}
\begin{equation}
\begin{split}
E_{1} &= 10 \hspace{1mm} \text{kPa}, \hspace{2mm} E_{2} = 30 \hspace{1mm} \text{kPa}, \hspace{2mm} {A} = 2, \hspace{2mm} {Y}_{2} = 1.1 \hspace{1mm} \text{cm},  \hspace{2mm} {D} = 0.003 \hspace{1mm} \text{cm}, \\
{Y}_{1} &= 0.115 \hspace{1mm} \text{cm}, \hspace{2mm}  (\text{for one-step variation--I}), \qquad {Y}_{1} = 0.145 \hspace{1mm} \text{cm}, \hspace{2mm}  (\text{for one-step variation--II}).
\label{eq:appenE1SV22c}
\end{split}
\end{equation}
\end{subequations}
\subsection{Multi-step variations in material properties: Interplay of necking and wrinkling behaviors} \label{appencircase3} 
The two-step variation in the Young modulus for the membrane is expressed as
\begin{subequations}
\begin{equation}
\begin{split}
E(R) &= E_{1}+E_{2}\Bigg(\frac{\exp{{A} \frac{(R - {Y}_{3})}{{D}_{1}}}-1}{\exp{{A} \frac{(R - {Y}_{3})}{{D}_{1}}}+1}\Bigg)-E_{2}\Bigg(\frac{\exp{{A}\frac{(R - {Y}_{4})}{{D}_{1}}}-1}{\exp{{A}\frac{(R - {Y}_{4})}{{D}_{1}}}+1}\Bigg) \\ &+ E_{2}\Bigg(\frac{\exp{{A} \frac{(R - {Y}_{5})}{{D}_{1}}}-1}{\exp{{A} \frac{(R - {Y}_{5})}{{D}_{1}}}+1}\Bigg)-E_{2}\Bigg(\frac{{N}_{1}\exp{{A}\frac{(R - {Y}_{6})}{{D}_{1}}}-1}{\exp{{A}\frac{(R - {Y}_{6})}{{D}_{1}}}+1}\Bigg),
\label{eq:appenE2SV23a}
\end{split}
\end{equation}
where
\begin{equation}
\begin{split}
E_{1} &= 10 \hspace{1mm} \text{kPa}, \hspace{2mm} E_{2} = 30 \hspace{1mm} \text{kPa}, \hspace{2mm} {A} = 2, \hspace{2mm} {D}_{1} = 0.0085 \hspace{1mm} \text{cm}, \hspace{2mm} {N}_{1} = 1.2,\\
{Y}_{3} &= 0.05 \hspace{1mm} \text{cm}, \hspace{1mm} {Y}_{4} = 0.1375 \hspace{1mm} \text{cm}, \hspace{1mm} 
{Y}_{5} = 0.225 \hspace{1mm} \text{cm}, \hspace{1mm} 
{Y}_{6} = 1.075 \hspace{1mm} \text{cm}.
\label{eq:appenE2SV23b}
\end{split}
\end{equation}
\end{subequations}
Similarly, the three-step variation in the Young modulus for the membrane is expressed as
\begin{subequations}
\begin{equation}
\begin{split}
E(R) &= E_{1}+E_{2}\Bigg(\frac{\exp{{A} \frac{(R - {Y}_{7})}{{D}_{2}}}-1}{\exp{{A} \frac{(R - {Y}_{7})}{{D}_{2}}}+1}\Bigg)-E_{2}\Bigg(\frac{\exp{{A}\frac{(R - {Y}_{8})}{{D}_{2}}}-1}{\exp{{A}\frac{(R - {Y}_{8})}{{D}_{2}}}+1}\Bigg) \\ &+ E_{2}\Bigg(\frac{\exp{{A} \frac{(R - {Y}_{9})}{{D}_{2}}}-1}{\exp{{A} \frac{(R - {Y}_{9})}{{D}_{2}}}+1}\Bigg)-E_{2}\Bigg(\frac{\exp{{A}\frac{(R - {Y}_{10})}{{D}_{2}}}-1}{\exp{{A}\frac{(R - {Y}_{10})}{{D}_{2}}}+1}\Bigg) \\ &+ E_{2}\Bigg(\frac{\exp{{A} \frac{(R - {Y}_{11})}{{D}_{2}}}-1}{\exp{{A} \frac{(R - {Y}_{11})}{{D}_{2}}}+1}\Bigg)-E_{2}\Bigg(\frac{{N}_{1}\exp{{A}\frac{(R - {Y}_{12})}{{D}_{2}}}-1}{\exp{{A}\frac{(R - {Y}_{12})}{{D}_{2}}}+1}\Bigg),
\label{eq:appenE3SV24a}
\end{split}
\end{equation}
where
\begin{equation}
\begin{split}
E_{1} &= 10 \hspace{1mm} \text{kPa}, \hspace{2mm} E_{2} = 30 \hspace{1mm} \text{kPa}, \hspace{2mm} {A} = 2, \hspace{2mm} {D}_{2} = 0.006 \hspace{1mm} \text{cm}, \hspace{2mm} {N}_{1} = 1.2, \hspace{2mm} {Y}_{7} = 0.085 \hspace{1mm} \text{cm}, \\
{Y}_{8} &= 0.1625 \hspace{1mm} \text{cm}, \hspace{1mm}
{Y}_{9} = 0.325 \hspace{1mm} \text{cm}, \hspace{1mm} 
{Y}_{10} = 0.5125 \hspace{1mm} \text{cm}, \hspace{1mm} 
{Y}_{11} = 0.7375 \hspace{1mm} \text{cm}, \hspace{1mm} 
{Y}_{12} = 1.0375 \hspace{1mm} \text{cm}.
\label{eq:appenE3SV24b}
\end{split}
\end{equation}
\end{subequations}
\section{Appendix: Square membranes}
\setcounter{equation}{0}
\renewcommand{\theequation}{C.\arabic{equation}}

\subsection{Distribution of the Young modulus in the auxetic square membranes}\label{appen-sq-YM}
The mathematical expression for the spatially varying Young's modulus of  a quarter of the square membrane (FG-SQ-1) reads
\begin{subequations}
\begin{equation} 
\begin{split}
E(X,Y) &= \pazocal{F}\left(\frac{L}{6} - X, \delta\right) \pazocal{F}\left(\sqrt{\big\lvert \left(\frac{L}{6}\right)^2 - X^2} \big \rvert - Y, \delta\right) E_{1} \\ & +
\pazocal{F}\left(\frac{L}{2} - X, \delta\right) \pazocal{F}\left(\frac{L}{2} - Y, \delta\right) E_{2},\\
\text{where} \hspace{1mm} \pazocal{F}\left(M, N\right) &= 0.5 + 0.9375 \left(\frac{M}{N}\right) - 0.625 \left(\frac{M}{N}\right)^3 + 0.1875\left(\frac{M}{N}\right)^5,\\
\text{and} \hspace{1mm} E_{1}  &= 0.27 \hspace{1mm} \text{MPa}, \hspace{2mm} E_{2}  = 0.03 \hspace{1mm} \text{MPa}, \hspace{2mm} \delta = 10^{-20}, \hspace{2mm} L = 1 \hspace{1mm} \text{m}, \hspace{2mm} 0 \leq X, Y \leq \frac{L}{2}.
\end{split}
\label{eq:sq-ym-fg1-26a}
\end{equation}
Here, the left bottom vertex of the quarter square membrane is located at ($X=0, Y=0$). The side length of the quarter square membrane in the undeformed configuration is $L/2$. The smoothed Heaviside function with a continuous second derivative is represented by $\pazocal{F}$, it is an inbuilt function in the \texttt{COMSOL} software. The function, $\lvert (\bullet) \rvert$, returns the absolute value of any variable, $(\bullet)$. \\
Similarly, for the full square membrane (FG-SQ-2) with side length $L$ in the undeformed configuration, the Young modulus can be mathematically expressed as
\begin{equation}
\begin{split}
E(X,Y) &= \pazocal{F} \left(X+\frac{2L_{1}}{3}, \delta\right) \pazocal{F}\left(\varepsilon - X, \delta\right) \pazocal{F}\left(3X + L_{1}-Y, \delta\right) \pazocal{F}\left(Y+\varepsilon, \delta\right)E_{1} \\ &+
\pazocal{F}\left(\frac{2L_{1}}{3} - X, \delta\right) \pazocal{F}\left(X-\varepsilon, \delta\right) \pazocal{F}\left(-3X + L_{1}-Y, \delta\right) \pazocal{F}\left(Y+\varepsilon, \delta\right)E_{1} \\ &+ 
\pazocal{F}\left(\frac{L}{2} - X, \delta\right) \pazocal{F}\left(\frac{L}{2} - Y, \delta\right)E_{2} 
\\ &+
\pazocal{F}\left(X+\frac{L_{1}}{3}, \delta\right) \pazocal{F}\left(-\frac{L_{1}}{6} - X, \delta\right) \pazocal{F}\left(-\frac{2L_{3}}{3}-Y, \delta\right) \pazocal{F}\left(Y+L_{2}, \delta\right)E_{3} 
\\ &+
\pazocal{F}\left(\frac{L_{1}}{3} - X, \delta\right) \pazocal{F}\left(X-\frac{L_{1}}{6}, \delta\right) \pazocal{F}\left(-\frac{2L_{3}}{3}-Y, \delta\right) \pazocal{F}\left(Y+L_{2}, \delta\right)E_{3}, \\
E_{1} = 0.6 & \hspace{1mm} \text{MPa}, \hspace{1mm} E_{2} = 0.03 \hspace{1mm} \text{MPa}, \hspace{1mm} E_{3} = 4 \hspace{1mm} \text{MPa},  L  = 1 \hspace{1mm} \text{m}, \hspace{1mm} L_{1} = \frac{2L}{7}, \hspace{1mm} L_{2} = \frac{L_{1}}{2}, \\
L_{3} &= \frac{L_{1}}{7}, \hspace{1mm} \delta = \varepsilon = 10^{-20}, \hspace{1mm} -\frac{L}{2} \leq X,Y \leq \frac{L}{2}.
\end{split}
\label{eq:sq-ym-fg2-26b}
\end{equation}
\end{subequations}



\bibliographystyle{elsarticle-harv-doichange}
\bibliography{cas-refs}

\end{document}